# Integrated high-performance error correction for continuous-variable quantum key distribution

Chuang Zhou,[1] Yang Li,[1,*] Li Ma,[1] Jie Yang,[1] Wei Huang,[1]
Ao Sun,[1] Heng Wang,[1] Yujie Luo,[1] Yong Li,[2] Ziyang Chen,[3]
Francis C. M. Lau,[4] Yichen Zhang,[5] Song Yu,[5] Hong Guo,[3] and Bingjie Xu,[1,‡]

[1]National Key Laboratory of Security Communication, Institute of Southwestern Communication, Chengdu 610041, China

[2]College of Computer Science, Chongqing University, Chongqing 400044, China

[3]State Key Laboratory of Advanced Optical Communication Systems and Networks, School of Electronics, and Center for Quantum Information Technology, Peking University, Beijing 100871, China

[4]Department of Electronic and Information Engineering, Hong Kong Polytechnic University, Hong Kong

[5]State Key Laboratory of Information Photonics and Optical Communications, Beijing University of Posts and Telecommunications, Beijing 100876, China

Abstract- An integrated error-correction scheme with high throughput, low frame errors rate (FER) and high reconciliation efficiency under low signal to noise ratio (SNR) is one of the major bottlenecks to realize high-performance and low-cost continuous variable quantum key distribution (CV-QKD). To solve this long-standing problem, a novel two-stage error correction method with limited precision that is suitable for integration given limited on-chip hardware resource while maintaining excellent decoding performance is proposed, and experimentally verified on a commercial FPGA. Compared to state-of-art results, the error-correction throughput can be improved more than one order of magnitude given FER<0.1 based on the proposed method, where 544.03 Mbps and 393.33 Mbps real-time error correction is achieved for typical 0.2 and 0.1 code rate, respectively. Besides, compared with traditional decoding method, the secure key rate (SKR) for CV-QKD under composable security framework can be improved by 140.09% and 122.03% by using the proposed two-stage decoding method for codes rate 0.2 and 0.1, which can support 32.70 Mbps and 5.66 Mbps real-time SKR under typical transmission distances of 25 km and 50 km, correspondingly. The record-breaking results paves the way for large-scale deployment of high-rate integrated CV-QKD systems in metropolitan quantum secure network.

## I. Introduction

Quantum key distribution (QKD) allows for secret key generation between two remote parties with provable information-theoretically security by leveraging the principles of quantum mechanics [1,2], which can be usually divided into two categories: discrete-variable quantum key distribution (DV-QKD) and continuous-variable quantum key distribution (CV-QKD) [3]. Due to the potential advantages of high secret key rate (SKR) within metropolitan areas and compatible with well compatibility with off-the-shelf optical devices working at room temperature, CV-QKD has attracted significant attention and made great breakthroughs in recent years [4–17] and more than hundreds of Mbps SKRs have been demonstrated [18–29].

However, the reported high-speed SKRs are only obtained by theoretical estimation or off-line post-processing, where the throughput of error correction is currently the major bottlenecks to realize practical high-performance CV-QKD. The existing CV-QKD repetition frequency can reach 10 GHz, while the error-correction throughput is only less than 100 Mbps. Furthermore, CV-QKD is currently advancing through photonic integration on its optical components to achieve high performance, cost reduction and compact size simultaneously, thereby facilitating the large-scale deployment [30-34]. However, to realize a complete integrated CV-QKD system, an integrated error-correction with high throughput, high reconciliation efficiency and low frame errors rate (FER) is strongly required.

---

[*] yishuihanly@pku.edu.cn
[‡] xbjpku@pku.edu.cn

To realize high reconciliation efficiency, well designed multi edge-type low density parity-check (MET-LDPC) codes with low code rate and long code length has been proven to be one of the most promising scheme for CV-QKD [35]. Recently, high-throughput error correction with MET-LDPC codes based on graphic processing unit (GPU) have been demonstrated [36-38]. However, the power consumption, volume, and cost of GPU limit its integration in CV-QKD system. Field programmable gate arrays (FPGA) is a kind of cost-effective large-scale programmable integrated devices with small volume, and the prototype verification on FPGA is an important step in the application specific integrated circuit (ASIC) design, which is currently the most favorite choice for integrated high-performance post-processing of CV-QKD. Therefore, FPGA-based MET-LDPC decoders for CV-QKD are studied very recently [39-41], where the currently reported error-correction throughput is very limited to maintain low FER. It is a great challenge to realize integrated error-correction with high-throughput and low FER based on FPGA.

The difficulty of a high-performance FPGA-based decoder for CV-QKD is mainly caused by two reasons. On the one hand, the on-chip hardware resource of FPGA is very limited, so that one must use fixed-point number with finite width (precision) to process raw data for a high-speed decoding, while the decoding throughput will be very limited if one uses floating-point number. On the other hand, the SNR for CV-QKD under typical transmission distance is much less than 0 dB, so the FER is strongly associated with the decoding accuracy. However, the limited precision for raw data decreases the decoding accuracy, leading to high FER after decoding, which will decrease the SKR of CV-QKD drastically [42-45]. For CV-QKD, traditional decoding method with limited precision confronts the problem of high FER. To our best knowledge, a decoder with limited precision and excellent performance under extremely low SNR has never been realized successfully for CV-QKD.

To solve the problem, a novel two-stage error-correction method with limited precision, while maintaining high throughput, low FER and high efficiency, is proposed for high-performance CV-QKD. The first stage is based on the Layered belief propagation (BP) decoding algorithm [40], which corrects most of bit errors and still leaves residual bit errors. In order to erase the residual bit errors, the effect of limited precision on FER is carefully studied, based on which a residual bit errors correction method is designed in the second stage to reduce FER. The proposed method is experimentally verified on a commercial FPGA. Compared to state-of-art results, the decoding throughput can be improved more than one order of magnitude given FER<0.1 based on the proposed method, where 544.03 Mbps and 393.33 Mbps real-time error-correction is achieved for typical 0.2 and 0.1 code rate, respectively. Besides, compared with traditional decoding method, the SKR for CV-QKD under composable security framework can be improved by 140.09% and 122.03% by using the proposed two-stage decoding method for codes rate 0.2 and 0.1, which can support 32.70 Mbps and 5.66 Mbps real-time SKR under typical transmission distances of 25 km and 50 km, correspondingly. The record-breaking results paves the way for large-scale deployment of high-rate integrated CV-QKD systems in metropolitan quantum secure network.

The rest of this paper is organized as follows. Section II introduces the problem of high FER under limited precision and the corresponding two-stage decoding method to solve the problem. Section III shows the experimental verification results and the corresponding real-rime SKR for a CV-QKD system that can be obtained by the proposed methods. Finally, Section IV draws a conclusion.

## II. Methods

The on-chip hardware resource of FPGA is very limited, so that one must use fixed-point number with finite precision to process raw data for high-throughput decoding. However, the limited precision will decrease the decoding accuracy based on the existing traditional decoding method, leading to high FER and low SKR, since the SNR of CV-QKD under typical transmission distance is extremely low. The problem of high FER with finite precision under low SNR is the bottleneck for integrated real-time CV-QKD system. To solve this problem, a novel two-stage error correction method with limited precision, which is suitable for integration given limited on-chip hardware resource while maintaining excellent decoding performance, is proposed.

### A. The problem of high FER with finite precision

In the reverse reconciliation scheme [46], Bob generates syndrome $s$ by $s = uH^T$, where vector $u$ is the raw keys of Bob and $H$ is the parity-check matrix of MET-LDPC codes. Then, Bob sends $s$ to Alice over classical channel. The data reconciliation of Alice generates data $R$, according to $R$ and $s$, Alice runs error-correction to obtain the exact same raw keys as Bob.

To verify the problem of high FER with finite precision, we experimentally demonstrate a FPGA based MET-LDPC decoder at typical rates 0.2 and 0.1 with traditionally used Layered BP decoding algorithm, where fixed-point number with finite width $w = 10$ (1 bit for the sign, 4 bits for the integer part and 5 bits for the fraction part) and $w = 12$ (1 bit for the sign, 4 bits for the integer part and 7 bits for the fraction part) are adopted to process raw data. For different SNRs, 100 data frames have been decoded, the FER vs SNR curves with different iterations for rates 0.2 and 0.1 are shown in Fig .1 and Fig. 2, respectively.

It shows that when using fixed-point number to design decoder, FER maintains a high level and increasing the number of decoding iterations or increasing the decoding SNR can not decrease FER effectively. However, when using floating-point number to design decoder, even though a small number of decoding iterations is adopted in the iterative decoding, FER decreases with the increase of SNR, and finally the FER is below 0.1.

It should be mentioned that the high FER is not restricted by decoding iterations and the FER can not be reduced to a reasonable value just by increasing the iterations. The finite width of the fixed-point number (limited decoding precision) results in high FER of the Layered BP decoding method.

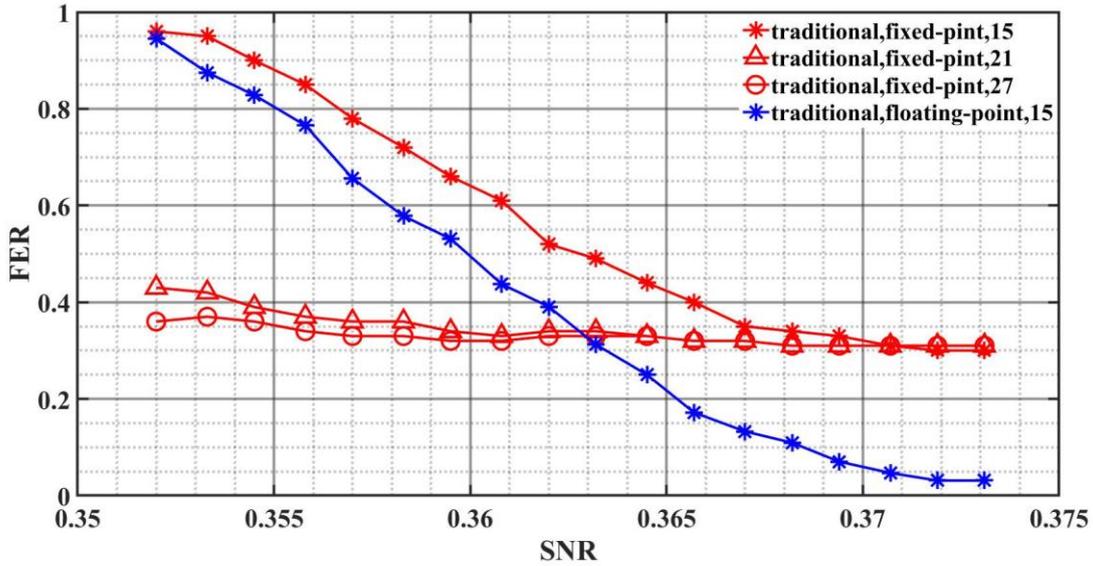

Fig.1 The FER vs SNR curve for a rate 0.2 LDPC code. The red curves show the result when fixed-point number with $w = 10$ is adopted in the iterative decoding with different iterations. The blue curve shows the result of floating-point number with the maximum number of iterations $t_{max} = 15$.

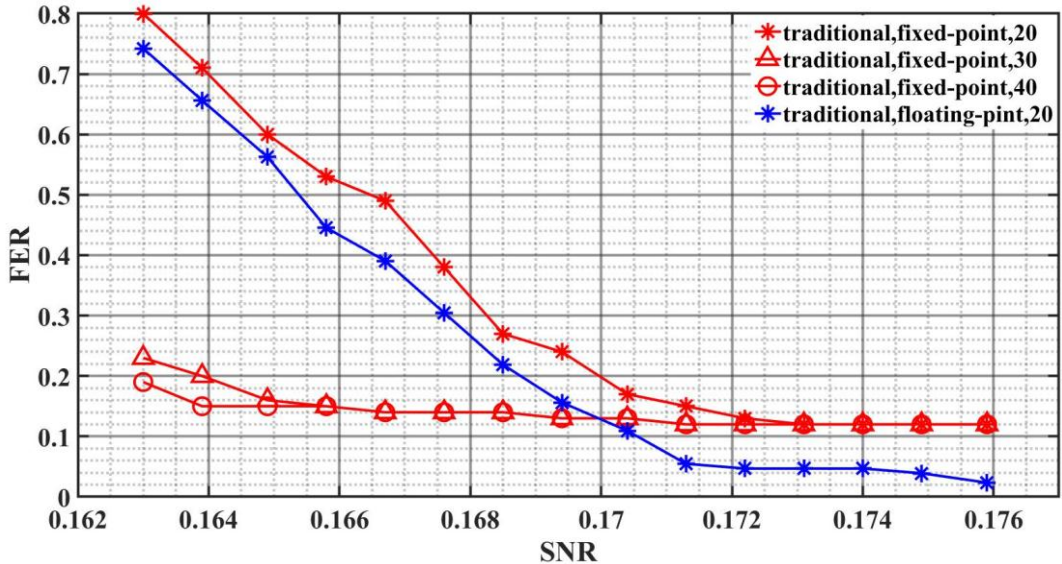

Fig.2 The FER vs SNR curve for a rate 0.1 LDPC code. The red curves show the result when fixed-point number with $w = 12$ is adopted in the iterative decoding with different iterations. The blue curve shows the result of floating-point number with the maximum number of iterations $t_{max} = 20$.

## B. Two-stage decoding method

In order to solve the problem of high FER, a novel two-stage decoding method with limited precision for CV-QKD is proposed. The first stage of the decoding method is based on the Layered BP decoding algorithm, which corrects most of bit errors and still leaves residual bit errors. The second stage of the decoding is to erase the residual bit errors without revealing any additional information.

In order to erase the residual bit errors, the effect of limited decoding precision on the FER is carefully studied. Define the reliability value as the absolute value of log likelihood ratio (LLR), and it is find that the reliability value has a deep correlation with residual bit errors. Take a rate 0.2 LDPC code with SNR = 0.3645 and $t_{max} = 15$ for example, four failed codewords are analyzed and the histogram of reliability value for the correct symbols and erroneous symbols after the first stage of the decoding method are drawn in Fig. 3(a) and Fig. 3(b). It is shown that the reliability value of erroneous symbols is obviously smaller than that of most correct symbols, which can be used to distinguish them efficiently. Hence, for a failed codeword, one can set a suitable threshold Δ. If the reliability value of a symbol is smaller than Δ, it is classified to the set of suspicious symbols, which belongs to an erroneous symbol with high probability.

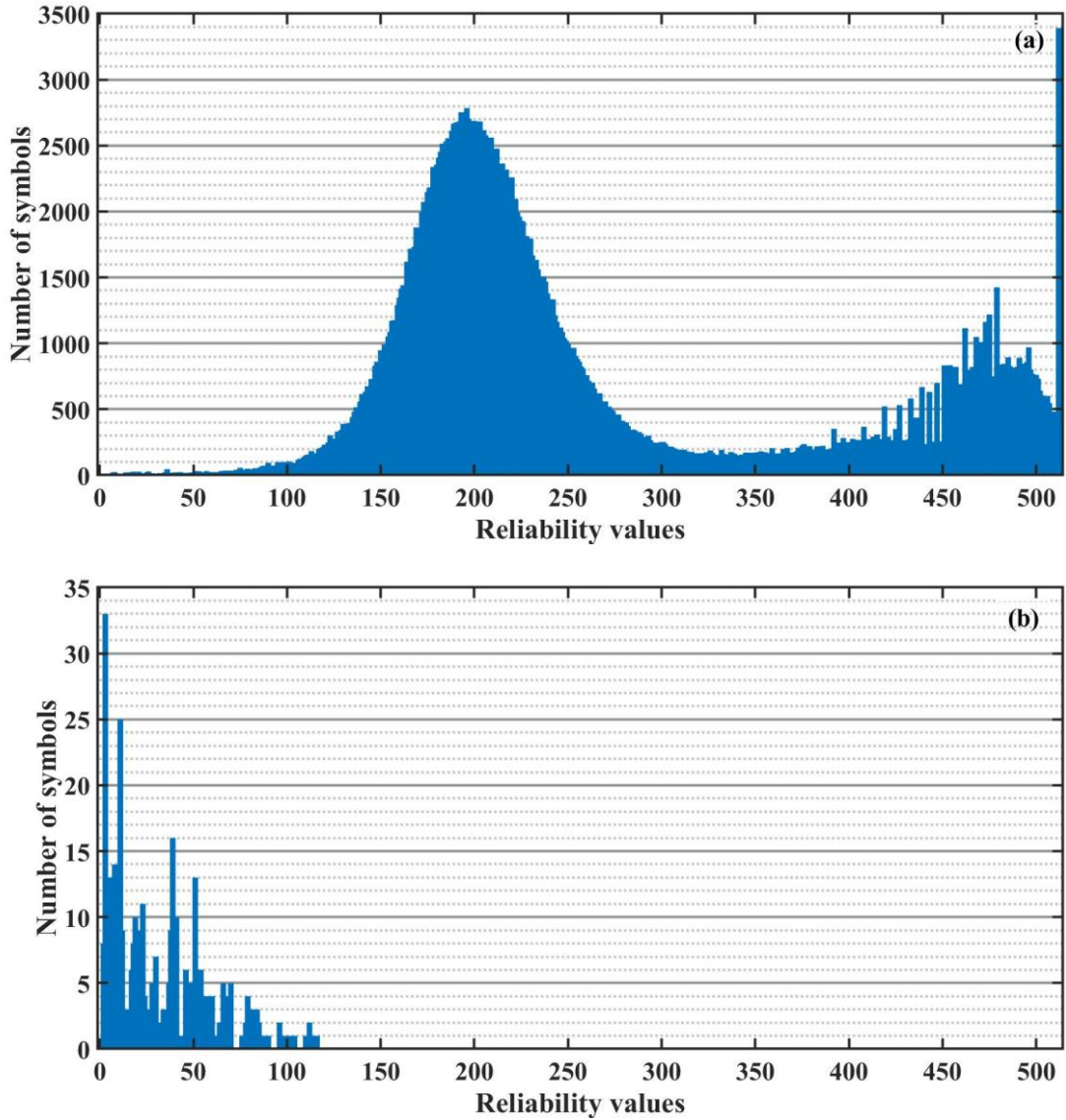

Fig.3 Histogram of decoded symbols after the first stage of the decoding method. (a): correct symbols; (b): erroneous symbols. Four failed codewords are collected at SNR = 0.3645 for a rate 0.2 LDPC code. The code length is set as 80000.

The choice of Δ has a great influence on the performance of the second stage of the decoding method. To obtain the most suitable value of Δ for different SNRs, analyzing the decoding results of each SNR and plotting the histogram as Fig. 3 are necessary. From the histogram, one can obtain which value of Δ can add all the erroneous symbols and the fewest correct symbols to the set of suspicious symbols. However, finding the optimal value of Δ for each SNR is complicated and it is not within the scope of this paper. For simplicity, fixed sub-optimal Δ is chosen for different SNRs in this paper.

Because of a few correct symbols are added to the set of suspicious symbols, just flipping decoding bits of these suspicious symbols can not erase the residual bit errors effectively. However, the syndrome $s$ can be used to judge whether the suspicious symbols are correct symbols or not. The residual bit errors correction method based on syndrome $s$ is described as follows [47]:

---

**The residual bit errors correction method**

---

(a) Generating sets $e$ and $\bar{e}$ based on the threshold $\Delta$.

$$\begin{cases} e = e \cup n, & |LLR_n| < \Delta \\ \bar{e} = \bar{e} \cup n, & |LLR_n| \geq \Delta \end{cases} \tag{1}$$

where $e$ and $\bar{e}$ contain the index of erroneous bits and correct bits that distinguished by the threshold $\Delta$. Matrix $H = \{h_n, n = 1, 2, 3, \ldots, N\}$ is the parity-check matrix of MET-LDPC codes, parity-check matrix $H_e = \{h_n, n \in e\}$ and $H_{\bar{e}} = \{h_n, n \in \bar{e}\}$.

(b) Computing syndrome $s_c$.

$$s_c = s \wedge s_{\bar{e}}, \tag{2}$$

where $s_{\bar{e}} = \hat{u}_{\bar{e}} H_{\bar{e}}^T$, $\hat{u}_{\bar{e}} = \{\hat{u}_n, n \in \bar{e}\}$ and $\hat{u}_n$ is the decode bit that generated by

$$\hat{u}_n = \begin{cases} 1, & LLR_n \leq 0 \\ 0, & LLR_n > 0 \end{cases}. \tag{3}$$

(c) If $H_e[m, n] = 1$, and the row-weight of the $m$'th row of matrix $H_e$ is 1, $\hat{u}_n = S_c[m]$ and one erroneous bit in $e$ is erased. Moving element $n$ from set $e$ to $\bar{e}$.

(d) Repeating step (b) to (c) until there are no rows of matrix $H_e$ with row-weight 1.

---

For the decoder of rate 0.2 LDPC codes, we set $\Delta = 165$, and 100 codewords are decoded for different SNRs as in Table I with the maximum number of decoding iterations $t_{max} = 15$. It shows that more than 75% of decoding failed codewords can be decoded successfully in the second stage of the decoding by using the residual bit errors correction method. Besides, the proposed two-stage decoding method needs relatively few iterations in the first stage to obtain the reliability values that can be completely classified by using $\Delta$. And then, the second stage using the residual bit errors correction method to decode successfully. The proposed two-stage decoding method can improve decoding throughput because of fewer iterations

Table I. The number of decoding failed codewords after the first stage and the number (percentage) of codewords that decoding successful in the second stage.

| SNR | 0.3595 | 0.3645 | 0.3694 | 0.3731 |
|---|---|---|---|---|
| Number of decoding failed codewords | 66 | 44 | 33 | 30 |
| Number (percentage) of decoding successful codewords | 50 (75.75%) | 39 (88.63) | 28 (84.84%) | 24 (80.00%) |

### III. Results and Analysis

## A. Decoding FER

In the second stage of the decoding, we have implemented the residual bit errors correction method on a commercial FPGA to improve the decoding performance of the first stage. The purple curves in Fig. 4 show the final decoding results of a rate 0.2 LDPC code with $t_{max}$ = 15 and 24. We can observe that the decoding FER of the proposed two-stage method with fixed-point numbers is very close to or even better than that with floating-point numbers, and much better than that with fixed-point numbers based on traditionally used Layered BP decoding method. Similarly, for the rate-0.1 MET-LDPC code, the decoding results of the proposed two-stage method is shown in Fig. 5, where the residual bit errors correction method can also significantly decrease FER.

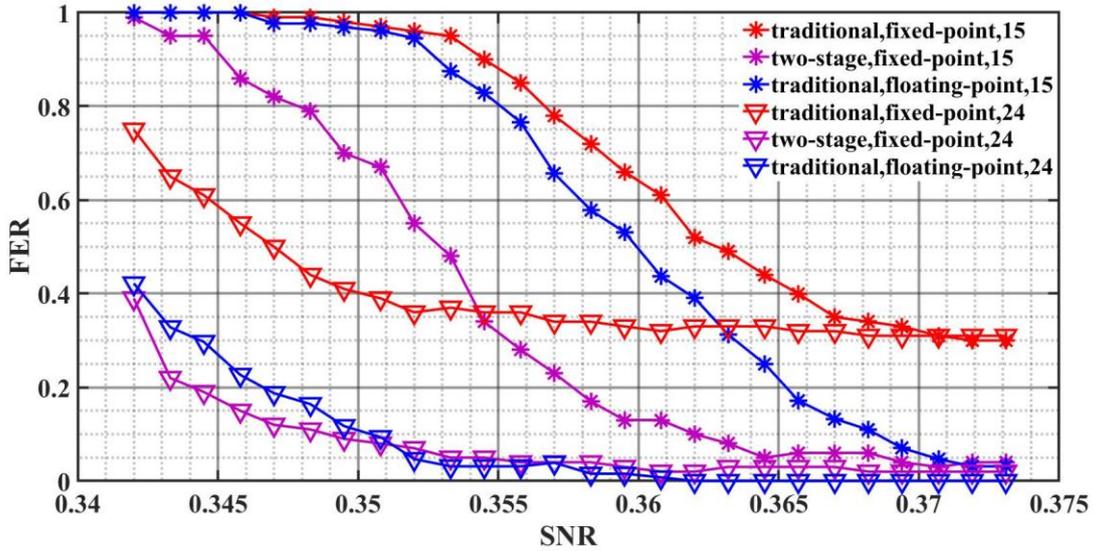

Fig.4 The FER vs SNR curves of the decoder for a rate-0.2 LDPC code. For each SNR, the decoding result of 100 codewords are collected under $t_{max}$ = 15 and 24, and $\Delta$ = 165. The red/purple curve shows the decoding result of this decoder with traditional/two-stage decoding method. The blue curve shows the decoding result of this code when floating-point number based on traditional decoding method is adopted in the iterative decoding.

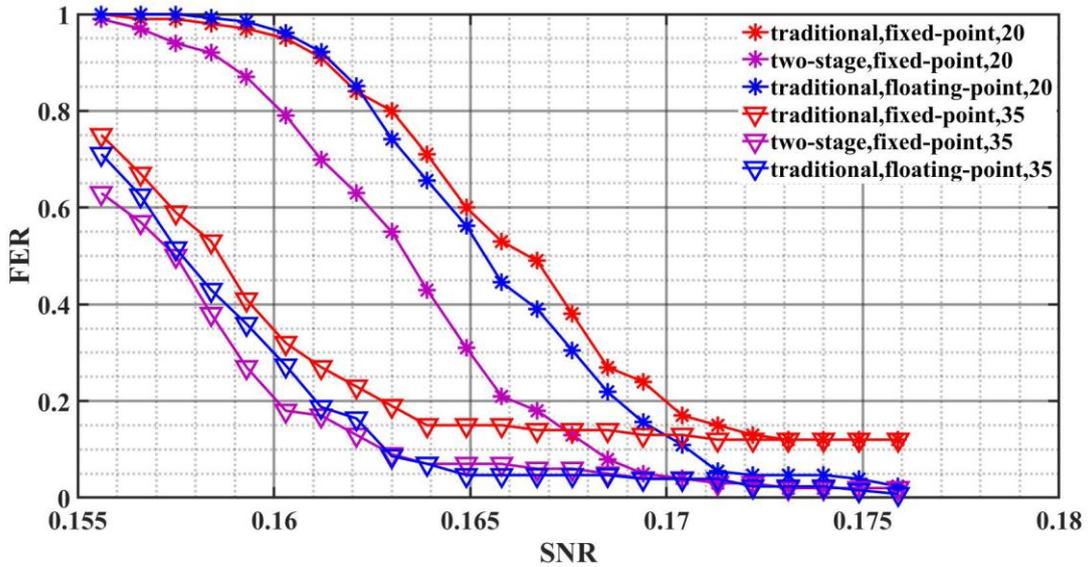

Fig.5 The FER vs SNR curves of the decoder for a rate-0.1 LDPC code. For each SNR, the decoding result of 100 codewords are collected under $t_{max}$ = 20 and 35, and $\Delta$ = 530. The red/purple curve shows the decoding result of this decoder with traditional/two-stage decoding method. The blue curve shows the decoding result of this code when floating-point number based on traditional decoding method is adopted in the iterative decoding.

## B. Decoding throughput

The proposed methods are implemented on FPGA as shown in Fig. 6, where module *sub_decoder* implements the function of the first stage and module *error_bits_erase* implements the function of the second stage. To make full use of the hardware resources and get higher decoding throughput, three *sub_decoder* modules and one *error_bits_erase* module are designed.

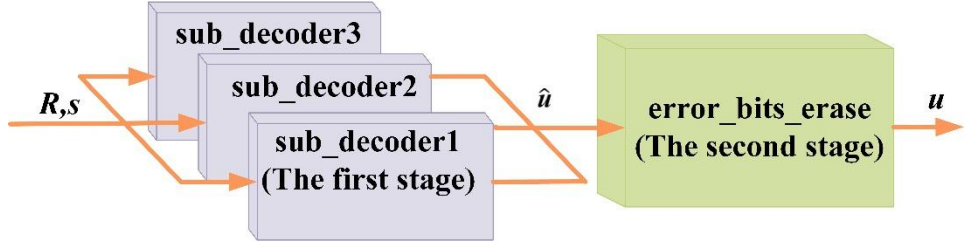

Fig. 6 The architecture of the proposed two-stage LDPC decoder

For the rate-0.2 (0.1) LDPC code, the consumption of clock cycles of the *error_bits_erase* module is approximately equal to 2.4 (3) decoding iterations. So, one *error_bits_erase* module can support the correction of residual bit errors of three *sub_decoder* modules when a pipeline structure is adopted to design the decoder with the maximum iterations are 15 and 35, which results in the second stage has no effect on the throughput of the first stage.

Table II. Performance of the implemented LDPC decoder and its comparison of other decoders designed for CV-QKD.

| Decoder | Device | Code rate | Data type | $t_{max}$ | SNR | FER | Decoding throughput (Mbps) | Real-time SKR (Mbps) |
|---|---|---|---|---|---|---|---|---|
| Wang. et al. [35] | GPU (NIVIDA TITAN XP) | 0.1 | floating-point | 100 | 0.16 | 0.055 | 30.39 | 0.57 |
| Li. et al. [36] | GPU (NIVIDA TITAN XP) | 0.1 | floating-point | 50 | 0.161 | 0.1797 | 72.86 | 1.15 |
| Yang. et al. [37] | GPU (NIVIDA GeForce RTX 3090) | 0.1 | floating-point | 100 | 0.16 | 0.093 | 65.5 | 1.17 |
| Yang. et al. [40] | FPGA XC7VX690T | 0.1 | fixed-point | 50 | 0.16 | 0.19 | 9.6 | 0.15 |
| This work | FPGA XC7VX690T | 0.1 | fixed-point | 35 | 0.16 | 0.18 | 226.37 | 3.71 |
| | | 0.1 | fixed-point | 20 | 0.1686 | 0.0858 | 393.33 | 5.66 |
| | | 0.2 | fixed-point | 15 | 0.3675 | 0.0343 | 544.03 | 32.70 |

The decoding throughput of a *sub_decoder* module can be estimated by

$$T = \frac{f_c \times N_{LDPC}}{D \times t_{max}}, \tag{4}$$

where $f_c$ is the frequency of clock on FPGA, $N_{LDPC}$ is code length, $D$ is the clock cycles used to finish one iteration decoding, and $t_{max}$ is the maximum number of decoding iterations.

Table II shows the decoding performance of FPGA-based LDPC decoder for rate 0.2 and 0.1 LDPC codes and compares them with other integrated decoders for CV-QKD. For the rate 0.1 LDPC decoder, because of using the proposed two-stage decoding method, fewer iterations ($t_{max} = 35$) is adopted to design the decoder, which improves the decoding throughput effectively and 226.37 Mbps decoding throughput is achieved. Compared with integrated decoders based on GPU [35-37], the decoding throughput improved more than 3 times with no loss of decoding efficiency. Compared with integrated decoders based on FPGA [40], the decoding throughput of rate 0.1 LDPC decoder in this work improved more than one order of magnitude under almost the same SNR and FER. When $t_{max} = 20$ is adopted to design decoder, the decoding throughput of rate 0.1 LDPC decoder can achieve 393.33 Mbps. For the rate 0.2 LDPC decoder, the decoding throughput in this work is up to 544.03 Mbps.

## C. Influence on composable finite-size secure key rates

SKR is the key indicator for CV-QKD systems. After parameter estimation, $K$ signals of each raw data block with size $N$ will be processed by error correction and privacy amplification. The composable finite-size secure key rates per pulse for no-switching CV-QKD protocol is given as [44,45],

$$SKR = \frac{K(1-FER)}{N}\left(\beta I_{AB} - \chi_{BE} - \frac{\Delta_{aep}}{\sqrt{K}} + \frac{\Theta}{K}\right), \tag{5}$$

where $\beta$ is the reconciliation efficiency, $I_{AB}$ is the mutual information between Alice and Bob, $\chi_{BE}$ is the mutual Information between Bob and Eve, and

$$\Delta_{aep} = 4 log_2(\sqrt{d}+2)\sqrt{log_2(\frac{18}{p_{ec}^2 \varepsilon_s^4})}, \tag{6}$$

$$\theta = log_2\left[p_{ec}\left(1 - \frac{\varepsilon_s^2}{3}\right)\right] + 2log_2\sqrt{2}\varepsilon_h, \tag{7}$$

with $d$ as the size of Alice's and Bob's effective alphabet after analog-to-digital conversion, $p_{ec} = 1 - FER$, $\varepsilon_h$ as a hashing parameter and $\varepsilon_s$ as a smoothing parameter.

As shown in [48], given known system parameters (e.g. channel transmittance, excess noise, detection efficiency and electronic noise), $I_{AB}$ and $\chi_{BE}$ can be described as the function of modulation variance $V_A$, that is $I_{AB} = f_{I_{AB}}(V_A)$ and $\chi_{BE} = f_{\chi_{BE}}(V_A)$. Besides, given fixed error correction codes and information reconciliation algorithm, the reconciliation efficiency can also be described as

$$\beta = \frac{R}{\frac{1}{2}log_2(1+SNR)} = f_\beta(V_A). \tag{8}$$

The purple curves in Fig. 4 and Fig. 5 show the relationship between FER and SNR, based on which one can get the fitting function of $f_{FER}(V_A)$ [48]. Finally, the SKR can be rewritten as

$$SKR = \frac{K(1-f_{FER}(V_A))}{N}\left(f_\beta(V_A)f_{I_{AB}}(V_A) - f_{\chi_{BE}}(V_A) - \frac{f_{\Delta_{aep}}(V_A)}{\sqrt{K}} + \frac{f_\theta(V_A)}{K}\right). \tag{9}$$

From Eq. (9), one can get the maximum SKR $K_{opt}$ by optimizing $V_A$ for known system parameters and error correction codes. Fig. 7(a) and Fig. 7(b) show the optimized SKR for the codes of rates 0.2 and 0.1, where the system parameters are chosen as excess noise $\varepsilon = 0.005$, detection electrical noise $vel = 0.041$, detection efficiency $\eta = 0.606$, channel transmittance $S = 10^{-\alpha L/10}$ with $\alpha = 0.2$ dB/km and $L$ as transmission distance. Compared with traditional decoding method, it is clearly shown that the SKR can be significantly improved by using the proposed two-stage method.

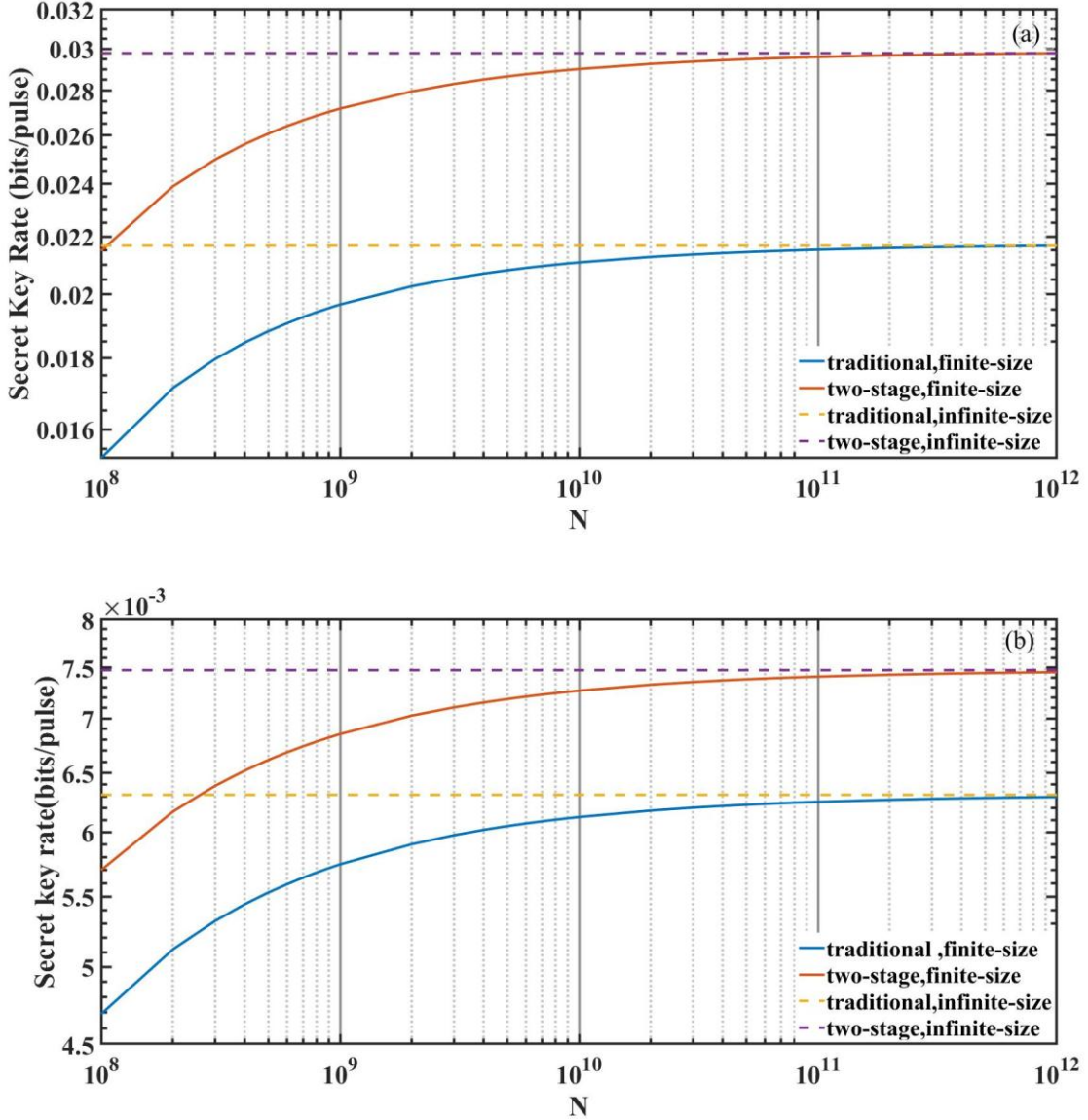

Fig. 7 Optimization results of SKR. (a): rate-0.2 LDPC code with $t_{max} = 15$ and $L = 25$ km. (b): rate-0.1 LDPC code with $t_{max} = 20$ and $L = 50$ km. Parameter $d = 32$, $\varepsilon_h = 10^{-10}$, $\varepsilon_s = 10^{-10}$ and $K/N = 0.5$.

More detailly, Table III shows the corresponding optimized SKR for transmission distance 25km and 50km. Parameter $G_K$ represents the improvements of SKR by using the two-stage method. Compared with using the traditional decoding method, the optimal SKR for rates 0.2 and 0.1 LDPC codes are improved up to 140.52% and 121.28% given $N = 10^8$ and 140.09% and 122.03% given $N = 10^{12}$ by using the proposed method.

Table III. The optimized SKR for $N = 10^8$ and $10^{12}$ with traditional decoding method and two-stage decoding methods.

| $R$ | $L$(km) | $N$ | method | FER | SNR | $K_{opt}$ | $G_K$ |
|---|---|---|---|---|---|---|---|
| 0.2 | 25 | $10^8$ | traditional | 0.2954 | 0.3715 | 0.0153 | 140.52% |
| | | | two-stage | 0.0896 | 0.3623 | 0.0215 | |
| | | $10^{12}$ | traditional | 0.3000 | 0.3699 | 0.0429 | 140.09% |
| | | | two-stage | 0.0343 | 0.3675 | 0.0601 | |
| 0.1 | 50 | $10^8$ | traditional | 0.1853 | 0.1702 | 0.0047 | 121.28% |
| | | | two-stage | 0.1083 | 0.1680 | 0.0057 | |
| | | $10^{12}$ | traditional | 0.1644 | 0.1708 | 0.0118 | 122.03% |
| | | | two-stage | 0.0858 | 0.1686 | 0.0144 | |

In a word, compared with traditional decoding results, by using this two-stage error-correction method, the FER for the decoder based on FPGA can be reduced obviously and the SKR can achieve a significant improvement. In fact, the real-time throughput of CV-QKD is mainly determined by the error correction throughput. By multiplying the optimal SKR and the throughput of the error correction, the method proposed in this paper can support 32.70 Mbps and 5.66 Mbps real-time SKR ($N = 10^{12}$) under transmission distance of 25 km and 50 km, correspondingly.

The last column of Table II shows the real-time SKR that the integrated decoders can support. The results in this work are much higher than other previously reported decoders. Besides, the SKR for rate-0.1 LDPC decoder with $t_{max} = 35$ are optimized by using Eq. (9) and the value of 0.0164 bits/pulse is achieved, which is higher than 0.0144 bits/pulse of the decoder with $t_{max} = 20$. However, the decoder with $t_{max} = 20$ can support higher real-time SKR because of much higher decoding throughput. So, how to choose the number of iterations can support the maximum generation of real-time SKR is an interesting problem and it is related to the design of the whole CV-QKD system.

## VI. Conclusion

In this paper, a novel two-stage decoding method with limited precision for integrated CV-QKD is proposed. Compared to state-of-art results, the decoding throughput can be improved more than an order of magnitude given FER<0.1 based on the proposed method, where 544.03 Mbps and 393.33 Mbps real-time error-correction is achieved for typical 0.2 and 0.1 code rate, respectively. Besides, compared with traditional decoding method, the secure key rates (SKR) for CV-QKD under composable security framework can be improved by 140.09% and 122.03% by using the proposed two-stage decoding method for codes rate 0.2 and 0.1, which can support 32.70 Mbps and 5.66 Mbps real-time SKR under typical transmission distances of 25 km and 50 km, correspondingly. The implementation results based on FPGA in this work also can be used in ASIC design, where the decoder can support much higher repetition frequency and realize real-time post-processing with much higher throughput. The record-breaking results paves the way for large-scale deployment of high-rate integrated CV-QKD systems in metropolitan quantum secure network.

## ACKNOWLEDGEMENT

We acknowledge financial support from the National Key Research and Development Program of China (Grant No. 2020YFA0309704), the National Natural Science Foundation of China (Grants No. 62101516, No. 62171418, No. 62201530, and No. 62301517), the Sichuan Science and Technology Program (Grants No. 2023ZYD0131, No. 2023JDRC0017, No. 2023YFG0143, No. 2022ZDZX0009 and No. 2021YJ0313), the Natural Science Foundation of Sichuan Province (Grants No. 2023NSFSC1387, No. 2023NSFSC0449, No. 2024NSFSC0470, and No.

2024NSFSC0454), the Basic Research Program of China (Grant No. JCKY2021210B059), the Equipment Advance Research Field Foundation (Grant No. 315067206), the National Key Laboratory of Security Communication Foundation (Grant No. 6142103042201, No. 6142103042301), Stability Program of National Key Laboratory of Security Communication (2023).

REFERENCES

[1] N. Gisin, G. Ribordy, W. Tittel, and H. Zbinden, "Quantum cryptography," *Rev. Mod. Phys.*, vol. 74, no. 1, pp.145-195, Mar. 2002.

[2] V. Scarani, H. Bechmann-Pasquinucci, N. J. Cerf, M. Dusek, Ń. Lutkenhaus, and M. Peev, "The security of practical quantum key distribution," Rev. Mod. Phys., vol. 81, no. 3, pp. 1301-1350, Sep. 2009.

[3] Y. Zhang, Y. Bian, Z. Li, S. Yu, and H. Guo, "Continuous-variable quantum key distribution system: Past, present, and future," *Appl. Phys. Rev.*, vol. 11, art. no. 011318. Mar. 2024.

[4] S. Pirandola, U. L. Andersen, L. Banchi, M. Berta, D. Bunandar, R. Colbeck, D. Englund, T. Gehring, C. Lupo, C. Ottaviani, J. L. Pereira, M. Razavi, J. Shamsul Shaari, M. Tomamichel, V. C. Usenko, G. Vallone, P. Villoresi, and P. Wallden, "Advances in quantum cryptography," *Adv. Opt. Photon.*, vol. 12, no. 4, pp. 1012-1236, Feb. 2020.

[5] S. Pirandola, "End-to-end capacities of a quantum communication network," *Commun. Phys.*, vol. 2, no. 1, art. no. 51, May 2019.

[6] A. Weerasinghe, M. Alhussein, A. Alderton, A. Wonfor, and R. Penty, "Practical, high-speed Gaussian coherent continuous variable quantum key distribution with real-time parameter monitoring, optimized slicing, and post-processed key distillation," *Sci. Rep.*, vol. 13, art. no. 21543, Dec. 2023.

[7] F. Grosshans, G. V. Assche, J. Wenger, R. Brouri, N. J. Cerf, and P. Grangier, "Quantum key distribution using gaussian-modulated coherent states," *Narture*, vol. 421, no. 6920, pp. 238-241, Jan. 2003.

[8] T. Matsuura, K. Maeda, T. Sasaki, and M. Koashi, "Finite-size security of continuous-variable quantum key distribution with digital signal processing," *Nat. Commun.*, vol. 12, no. 1, art. no. 252, Jan. 2021.

[9] Weedbrook, A. M. Lance, and W. P. Browen, "Quantum cryptography without switching," *Phys. Rev. Lett.*, vol. 93, no. 17, art. no. 170504, Oct. 2004.

[10] Weedbrook, S. Pirandola, R. Garcia-Patron, N. J. Cerf, T. C. Ralph, J. H. Shapiro, and S. Lloyd, "Gaussian quantum information," *Rev. Mod. Phys.*, vol. 84, no. 2, pp. 631-669, May 2012.

[11] P. Jouguet, S. Kunz-Jacques, A. Leverrier, P. Grangier, and E. Diamanti, "Experimental demonstration of long-distance continuous-variable quantum key distribution," *Nat. Photonics*, vol. 7, no. 5, pp. 378-381, Apr. 2013.

[12] Z. Chen, X. Wang, S. Yu, Z. Li and H. Guo, "Digital Quantum Key Distribution with Continuous-Mode Formalism," *Npj Quantum Inf.*, vol. 9, no. 1, art. no. 28, Mar. 2023.

[13] S. Ghorai, P. Grangier, E. Diamnti, and A. Leverrier, "Asymptotic security of continuous-variable quantum key distribution with a discrete modulation," *Phys. Rev. X*, vol. 9, no. 2, p. 021059, Jun. 2019.

[14] A. Denys, P. Brrown, and A. Leverrier, "Explicit asymptotic secret key rate of continuous-variable quantum key distribution with an arbitrary modulation," *Quantum*, vol. 5, no. 1, p.

540, Sep. 2021.

[15] W. B. Liu, C. L. Li, Y. M. Xie, C. X. Weng, J. Gu, X. Y. Cao, Y. S. Lu, B. H. Li, H. L. Yin, and Z. B. Chen, "Homodyne detection quadrature phase shift keying continuous-variable quantum key distribution with high excess noise tolerance," *PRX Quantum*, vol. 2, no. 4, p. 040334, Nov. 2021.

[16] B. Qi, P. Lougovski, R. Pooser, W. Grice, and M. Bobrek, "Generating the local oscillator "locally" in continuous-variable quantum key distribution based on coherent detection," *Phys. Rev. X*, vol. 5, no. 4, p. 041009, Oct. 2015.

[17] S. J. Johnson, A. M. Lance, L. Ong, M. Shirvanimoghaddam, T C Ralph and T. Symul, "On the problem of non-zero word error rates for fixed-rate error correction codes in continuous variable quantum key dstribution," *New J. Phys.*, vol. 10, no. 2, p. 023003, Feb. 2017.

[18] Y. Tian, P. Wang, J. Q. Liu, S. N. Du, W. Y. Liu, Z. G. Lu, X. Y. Wang, and Y. M. Li, "Experimental demonstration of continuous-variable measurement-device-independent quantum key distribution over optical fiber," *Optical*, vol. 9, no. 5, p.492, May 2022.

[19] N. Jain, H. Chin, H. Mani, C. Lupo, D. S. Nikolic, A. Kordts, S. Pirandola, T. B. Pedersen, M. Kolb, B. Ó, C. Pacher, T. Gehring, and U. L. Andersen, "Practical continuous-variable quantum key distribution with composable security," *Nat. Commun.*, vol. 13, no. 1, art. no. 4740, Aug. 2022.

[20] Y. Tian, Z. Yu, S. S. Liu, P. Wang, Z. G. Lu, X. Y. Wang, and Y. M. Li, "High-performance long-distance discrete-modulation continuous-variable quantum key distribution," *Opt. Let.*, vol. 48, no. 11, pp.2953-2956, May 2023.

[21] Y. D. Pi, H. Wang, Y. Shao, Y. Li, J. Yang, Y. C. Zhang, W. Huang, and B. J. Xu, "Sub-Mbps key-rate continuous-variable quantum key distribution with local local oscillator over 100-km fiber," *Opt. Let.*, vol. 48, no. 7, pp. 1766-1769, Mar. 2023.

[22] Y. Zhang, Z. Chen, S. Pirandola, X. Wang, C, Zhou, B. Chu, Y. Zhao, B. Xu, S. Yu, and H. Guo, "Long-distance continuous-variable quantum key distribution over 202.81 km of fiber," *Phys. Rev. Lett.*, vol. 125, no. 1, art. no. 010502, Jun. 2020.

[23] H. Wang, Y. Pi, W. Huang, Y. Li, Y. Shao, J. Yang, J. Liu, C. Zhang, Y. Zhang, and B. Xu, "High-speed Gaussian-modulated continuous-variable quantum key distribution with a local local oscillator based on pilot-tone-assisted phase compensation," *Opt. Express*, vol. 28, no. 22, p. 32882, Oct. 2020.

[24] H. Wang, Y. Li, Y. Pi, Y. Pan, Y, Shao, L. Ma, Y, Zhang, J. Yang, T. Zhang, W. Huang, and B. Xu, "Sub-Gbps key rate four-state continuous-variable quantum key distribution within metropolitan area," *Commun. Phys.*, vol. 5, no. 1, art. no. 162, Jun. 2022.

[25] S. Ren, S. Yang, A. Wonfor, I. White, and R. Penty, "Demonstration of high-speed and low-complexity continuous variable quantum key distribution system with local local oscillator," *Sci. Rep.*, vol. 11, no. 1, art. no. 9454, May 2021.

[26] Roumestan, A. Ghazisaeidi, J. Renaudier, L. T. Vidarte, E. Diamanti and P. Grangier, "High-rate continuous variable quantum key distribution based on probabilistically shaped 64 and 256-QAM," in *2021 European Conference on Optical Communication (ECOC)*, Bordeaux, France, 2021, pp. 1-4.

[27] Y. Pan, H. Wang, Y. Shao, Y. Pi, Y. Li, B. Liu, W. Huang, and B. Xu, "Experimental demonstration of high-rate discrete-modulated continuous-variable quantum key distribution system," *Opt. Lett.*, vol. 47, no. 13, pp. 3307-3310, Jul. 2022.

[28] T. Wang, P. Huang, Y. Zhou, W. Liu, H. Ma, S. Wang, and G. Zeng, "High key rate continuous-variable quantum key distribution with a real local oscillator," *Opt. Express*, vol. 26, no. 3, pp. 2794-2806, Feb. 2018.

[29] T. A. Eriksson, R. S. Luis, B. J. Puttnam, G. Rademacher, M. Fujiwara, Y. Awaji, H. Furukawa, N. Wada, M. Takeoka, and M. Sasaki, "Wavelength division multiplexing of 194 continuous variable quantum key distribution channels," *J. Lightwave Technol.*, vol. 38, no. 8, pp. 2214-2218, Apr. 2020.

[30] G. Zhang, J. Y. Haw, H. Cai, F. Xu, S. M. Assad, J. F. Fitzsimons, X. Zhou, Y. Zhang, S. Yu, J. Wu, W. Ser, L. C. Kwek, and A. Q. Liu, "An integrated silicon photonic chip platform for continuous variable quantum key distribution," *Nat. photonics*, vol. 13, pp. 839-842, Aug. 2019.

[31] Y. M. Bian, Y. Li, X. S. Xu, T. Zhang, Y. Pan, W. Huang, S. Yu, L. Zhang, Y. C. Zhang, and B. J. Xu, "Highly stable power control for chip-based continuous-variable quantum distribution system," *Opt. Let.*, vol. 49, no. 9, pp. 2521-2524, May. 2024.

[32] Y. M. Bian, Y. Pan, X. S. Xu, L. Zhao, Y. Li, W. Huang, L. Zhang, S. Yu, Y. C. Zhang, B. J. Xu, "Continuous variable quantum key distribution over 28.6 km fiber with an integrated silicon photonic receiver chip," *Appl. phys. Lett.*, vol. 124, art. no. 174001, Apr. 2024.

[33] J. Wang, F. Sciarrino, A. Laing, and M. G. Thompson, "Integrated photonic quantum technologies," *Nat. photonics*, vol. 14, pp. 273-284, Oct. 2020.

[34] W. Luo, L. Cao, Y. Shi, et al. "Recent progress in quantum photonics chips for quantum communication and internet," *Light Sci. Appl.*, vol. 12, art. no. 175, Jul. 2023.

[35] S. Jeong, H. Jung, and J. Ha, "Rate-compatible multi-edge type low-density parity-check code ensembles for continuous variable quantum key distribution systems," *Npj Quantum Inf.*, vol. 8, no. 1, art. no. 6, Jan. 2022.

[36] X. Wang, Y. Zhang, S. Yu, and H. Guo, "High speed error correction for continuous-variable quantum key distribution with multi-edge type LDPC code," *Sci. Rep.*, vol. 8, no. 1, art. no. 10543, Jul. 2018.

[37] Y. Li, X. Zhang, Y. Li, B. Xu, L. Ma, J. Yang, and W. Huang, "High-throughput GPU layered decoder of quasi-cyclic multi-edge type low density parity check codes in continuous-variable quantum key distribution systems," *Sci. Rep.*, vol. 10, no. 1, art. no. 14561, Sep. 2020.

[38] H. Z. Yang, S. S. Liu, S. S. Yang, Z. G. Lu, Y. X. Li, and Y. M. Li, "High efficiency rate-adaptive reconciliation in continuous–variable quantum key distribution," *Phys. Rev. A*, vol. 109, no. 1, p. 012604, Jan. 2024.

[39] S. S. Yang, Z. G. Lu, and Y. M. Li, "High-speed post-processing in continuous-variable quantum key distribution based on FPGA implementation," *J. Lightwave Technol.*, vol. 38, no. 15, pp. 3935-3941, Aug. 2020.

[40] S. S. Yang, J. Q. Liu, Z. G. Lu, Z. L. Bai, X. Y. Wang and Y. M. Li, "An FPGA-based LDPC decoder with ultra-long codes for continuous-variable quantum key distribution," *IEEE Access*, vol. 9, pp. 47687-47697, 2021.

[41] S. S. Yang, Z. L. Yan, Q. Lu, H. Z. Yang, Z. G. Lu, X. Y. Miao, and Y. M. Li, "Hardware design and implementation of high-speed multidimensional reconciliation sender module in continuous-variable quantum key distribution," *Quantum Inf. Process.*, vol. 22, no. 10, pp. 1-14, Oct. 2023.

[42] S. Pirandola, R. Laurenza, C. Ottaviani, and L. Banchi, "Fundamental limits of repeaterless quantum communications," *Nat. Commun.*, vol. 8, no. 1, art. no. 15043, Apr. 2017.

[43] C. Lupo, C. Ottaviani, P. Papanastasiou, and S. Pirandola, "Continuous-variable measurement-device-independent quantum key distribution: Composable security against coherent attacks," *Phys. Rev. A*, vol. 97, no. 5, p. 052327, May 2018.

[44] S. Pirandola, "Limits and security of free-space quantum communications," *Phys. Rev. Research*, vol.3, no. 1, p. 013279, Mar. 2021.

[45] S. Pirandola, "Composable security for continuous variable quantum key distribution: Trust levels and practical key rates in wired and wireless networks," *Phys. Rev. Research*, vol. 3, no. 4, p. 043014, Oct. 2021.

[46] A. Leverrier, R. Alleaume, J. Boutros, G. Zemor and P. Grangier, "Multidimensional reconciliation for a continuous-variable quantum key distribution," *Phys. Rev. A*, vol. 77, no. 4, p. 042325, Apr. 2008

[47] H. Cui, J. Lin and Z. Wang, "An efficient post-processor for lowering the error floor of LDPC codes," *IEEE Transactions on Circuits and Systems II: Express Briefs*, vol. 66, no. 3, pp. 397-401, Mar. 2019.

[48] L. Ma, Y. Li, T. Zhang, Y. Shao, J. L. Liu, Y. J. Luo, H. Wang, W. Huang, F. Fan, and et al. "Practical continuous-variable quantum key distribution with feasible optimization parameters," *Sci. China Inf.*, vol. 66, no. 18, art. no. 180507, Jun. 2023.